\documentclass{article}
\usepackage[utf8]{inputenc}
\usepackage{amsmath, amssymb, amsthm}
\usepackage{graphicx}
\usepackage{tikz-cd}
\usepackage{booktabs}
\usepackage{longtable}
\usepackage{hyperref}
\usepackage{fullpage}

\newtheorem{theorem}{Theorem}
\newtheorem{definition}{Definition}

\title{A Sheaf Framework for Strategic Multi-Agent Systems:\\ From Consensus to Nash Equilibria}
\author{Manuel Hernandez \and Eduardo S\'anchez-Soto}
\date{May 30, 2026}

\begin{document}

\maketitle

\begin{abstract}
The coordination of heterogeneous autonomous agents in dynamic,
adversarial environments requires simultaneous satisfaction of
geometric constraints, logical consistency, temporal reasoning, and
strategic optimization. Existing sheaf- and topos-theoretic frameworks
provide powerful tools for geometric consensus, knowledge alignment,
and causal planning, but lack explicit models for value, reward, and
strategic choice. This report presents a unified categorical framework
that integrates event calculus, SCEL-like ensemble formation, and
game-theoretic reward structures into a single Grothendieck topos of
time-space histories. We introduce the notion of a \emph{game sheaf}
whose stalks contain utility functions and policy distributions, and
restriction maps encode both parallel transport and best-response
dynamics. We prove that Nash equilibria correspond to global sections
of a derived best-response correspondence sheaf, while cohomological
obstructions classify failures of strategic consistency. A detailed
case study of an immunological ``bastion defense''
scenario---heterogeneous agents forming attack/defense ensembles under
resource constraints---demonstrates the framework's
expressiveness. This synthesis provides a rigorous foundation for
verifiable, autonomic, and economically rational multi-agent systems.
\end{abstract}

\section{The Three Headaches of Swarm Coordination}
Modern multi-agent systems (MAS) operating in open, adversarial
environments—such as autonomous defense swarms, robotic
search-and-rescue teams, or distributed sensor networks—face three
fundamental challenges:
    a) \textbf{Geometric coordination}: agents must align their physical motions (poses, velocities) under non-holonomic constraints, often without a global coordinate system; b)
    \textbf{Logical and temporal consistency}: agents must maintain coherent beliefs, plan actions over time, and repair knowledge after unobserved interventions; and, c)
  \textbf{Strategic optimization}: agents must make decisions that maximize individual or collective utility under scarcity, uncertainty, and potential conflict of interest.

Recent advances using sheaf theory, Cartan geometry, and topos
semantics have addressed the first two challenges using cellular
sheaves \cite{robinson2014, curry2014} and interval temporal planning
\cite{hernandez2026sheaf, hernandez2026toposCausal}. However, none of
these approaches incorporates value or strategic choice. Agents in
real-world scenarios must trade off costs (energy, time, risk) against
rewards (territory, resources, mission success). They form ensembles
not only based on predicates but also on expected utility. They may
cooperate, compete, or defect. This report fills this gap by extending
the unified topos with a game-theoretic layer.

With this paper, we propose:
\begin{itemize}
    \item A formal integration of event calculus into the temporal site, so that rewards can be updated as actions happen.
    \item A \emph{game sheaf} whose global sections correspond to Nash equilibria, and where cohomology tells us whether a consistent strategy exists.
    \item A hybrid dynamics that blends sheaf Laplacian diffusion (for consensus) with gradient ascent on expected rewards.
    \item A running example: an immunological bastion defense, where scouts, artillery, and logistics units must cooperate to protect a high-value asset.
\end{itemize}

\section{A Brief Refresher: Sheaves, Topoi, and Why You Should Care}
We recall the three base frameworks; details can be found in the original references \cite{hernandez2026sheaf, hernandez2026explanation, hernandez2026toposVerifiable}.

\subsection{Geometric consensus via cellular sheaves}
A communication graph $G=(V,E)$ is equipped with a cellular sheaf $\mathcal{C}$: each vertex $v$ has a stalk $\mathcal{C}(v)=\mathcal{K}_v\times\mathcal{M}_v$ (knowledge and motor), each edge $e=\{u,v\}$ has a stalk $\mathcal{C}(e)$ (interaction space), and restriction maps $\mathcal{C}_{v\subseteq e}$ given by parallel transport (Cartan connection). The sheaf Laplacian drives consensus:
\[
\partial_t \phi = -L_{\mathcal{C}}\phi,\qquad (L_{\mathcal{C}}\phi)_v = \sum_{e\ni v} \mathcal{C}_{v\subseteq e}^*(\mathcal{C}_{v\subseteq e}\phi_v - \mathcal{C}_{u\subseteq e}\phi_u).
\]
Global sections $H^0(G;\mathcal{C})$ correspond to perfect consensus; obstructions are measured by $H^1(G;\mathcal{C})$.

\subsection{Temporal planning and event calculus}
Time is a poset category $\mathcal{T}$ of closed intervals with inclusions. A \emph{plan sheaf} on $\mathcal{T}$ assigns to each interval the set of consistent histories \cite{hernandez2026sheaf}. Actions are natural transformations. Event calculus axioms (inertia, causal consistency) become sheaf conditions \cite{hernandez2026toposCausal}.

\subsection{The product topos}
The combined site $\mathcal{S}=\mathcal{T}\times G$ (product category) with product topology yields the topos $\mathbf{Sh}(\mathcal{S})$. Objects are pairs $(I,v)$ with $I\in\mathcal{T}, v\in V(G)$. A sheaf on $\mathcal{S}$ assigns stalks containing both temporal histories and geometric/knowledge states.

\begin{figure}[h]
\centering
\begin{tikzcd}
\mathcal{T}\;(\text{time intervals}) \arrow[dr, "\text{product}"] &   \\
& \mathcal{T}\times G \arrow[d] \\
G\;(\text{comm. graph}) \arrow[ur, "\text{product}"] & \text{Sheaf } \mathcal{F} \text{ on } \mathcal{T}\times G
\end{tikzcd}
\caption{The product site combines temporal intervals and graph vertices. A sheaf on this site encodes both time-dependent and agent-dependent data.}
\label{fig:product}
\end{figure}
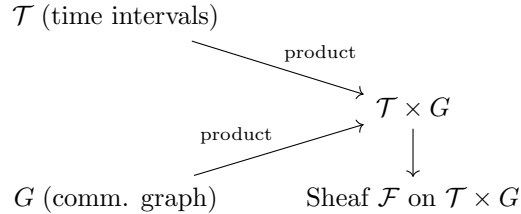

\section{Adding Carrots and Sticks: Rewards and Event Calculus}
\subsection{Reward sheaf}
A reward is a real-valued fluent. Define the \emph{reward sheaf} $\mathcal{R}$ on $\mathcal{S}$:
\begin{itemize}
    \item For each $(I,v)$, $\mathcal{R}(I,v)=\mathbb{R}^{m_v}$ (vector of reward components, e.g., health, ammo, score).
    \item Temporal restriction: if no action in $I\setminus I'$, $\rho_{I,I'}(r)=r$; otherwise add accumulated reward.
    \item Spatial restriction: for edge $e=(u,v)$, $\mathcal{R}_{v\subseteq e}:\mathcal{R}(v)\to\mathcal{R}(e)$ aligns reward estimates (e.g., average or max).
\end{itemize}
The total utility for agent $v$ over interval $I$ is $U_v(I)=w_v\cdot r_v(I)$.

\subsection{Event calculus as sheaf axioms}
Following \cite{hernandez2026toposCausal}, the sheaf conditions encode:
\begin{enumerate}
    \item \textbf{Inertia}: If no action affecting fluent $\phi$ occurs in $I$, then $\phi$ is constant on all subintervals.
    \item \textbf{Causal consistency}: If action $a$ initiates $\phi$ at time $t$, then any interval containing $t$ must have $\phi$ true after $t$ unless terminated.
\end{enumerate}
These are implemented as a sheaf of temporal structures.

\section{Game of Sheaves: When Agents Get Strategic}
\subsection{Definition of a game sheaf}
A \emph{game sheaf} $\mathcal{G}$ over $\mathcal{S}$ consists of:
\begin{itemize}
    \item A \emph{utility sheaf} $\mathcal{U}$: for each $(I,v)$, $\mathcal{U}(I,v)=\mathbb{R}^{n_v}$ (payoff vector for pure strategy profiles).
    \item A \emph{strategy sheaf} $\Sigma$: for each $(I,v)$, $\Sigma(I,v)$ is a compact convex set (mixed strategies).
    \item A \emph{reward sheaf} $\mathcal{R}$ with bilinear pairing $\langle\cdot,\cdot\rangle:\Sigma(I,v)\times\mathcal{U}(I,v)\to\mathbb{R}$.
    \item Restriction maps $\Sigma_{v\subseteq e}$ that align strategy profiles across edges.
\end{itemize}
A global section of $\Sigma$ is a family $\sigma=(\sigma_{I,v})$ such that for each edge $e$, $\Sigma_{v\subseteq e}(\sigma_{I,v})=\Sigma_{u\subseteq e}(\sigma_{I,u})$ — strategies are consistent across the communication graph. This links our model directly to compositional game theory \cite{hedges2018compositional} and compositional modeling of network games \cite{dilavore2021compositional}.

\subsection{Nash equilibrium as a sheaf condition}
\begin{definition}
A global section $\sigma$ of $\Sigma$ is a \emph{sheaf-theoretic Nash equilibrium} if for every agent $v$ and every alternative $\tau_v\in\Sigma(I,v)$,
\[
U_v(\tau_v,\sigma_{-v})\le U_v(\sigma_v,\sigma_{-v}),
\]
where utility is computed from $\mathcal{U}(I,v)$ using restriction maps to align with neighbors.
\end{definition}

\begin{theorem}
The set of sheaf-theoretic Nash equilibria corresponds to the global sections of the \emph{best-response sheaf} $\mathcal{B}$, defined stalkwise by
\[
\mathcal{B}(I,v)=\{\sigma_v\in\Sigma(I,v)\mid \sigma_v\in BR_v(\sigma_{-v})\}.
\]
A global section exists if and only if $H^1(\mathcal{S};\mathcal{B})=0$, where $\mathcal{B}$ is regarded as a sheaf of sets.
\end{theorem}
This theorem connects strategic equilibrium to sheaf cohomology: non-vanishing $H^1$ indicates that local best-responses cannot be glued into a global equilibrium — a \emph{strategic obstruction}.

\begin{figure}[h]
\centering
\begin{tikzcd}
& \mathcal{B}(U) \arrow[dr] \arrow[dd] & \\
\mathcal{B}(U\cap V) \arrow[ur] \arrow[dr] & & \mathcal{B}(V) \arrow[dd] \\
& \mathcal{B}(U\cap V) \arrow[ur] & \\
\mathcal{B}(U) \arrow[rr] & & \mathcal{B}(V)
\end{tikzcd}
\caption{The gluing condition for best-response sheaf. Local equilibria on overlapping covers must agree; otherwise $H^1$ obstructs a global Nash equilibrium.}
\label{fig:bestresponse}
\end{figure}
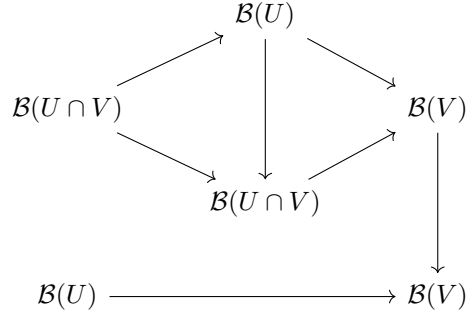

\subsection{Combined dynamics: diffusion + gradient ascent}
We propose a hybrid update rule for each agent $v$ at interval $I$:
\[
\phi_v^{(t+1)} = \phi_v^{(t)} - \alpha L_{\mathcal{C}}\phi_v^{(t)} + \beta \nabla_{\phi_v}U_v(\phi^{(t)}),
\]
where $\phi_v$ now includes geometric/knowledge components and strategy parameters, $L_{\mathcal{C}}$ is the sheaf Laplacian of the combined sheaf $\mathcal{C}_{\text{total}}=\mathcal{C}_{\text{geom}}\times\mathcal{C}_{\text{know}}\times\Sigma$, and $\nabla U_v$ is the gradient of agent $v$'s utility with respect to its own strategic variables. Under convexity assumptions (e.g., potential game plus positive semidefinite Laplacian), this dynamics converges to a point that is both a global section (consensus) and a Nash equilibrium.

\section{The Bastion Defense: An Immunological Parable}

\emph{Collaborative work in action.} We consider a heterogeneous swarm defending a bastion (a high-value asset) against an external attacker. Agents belong to three types:
\begin{itemize}
    \item \textbf{Scouts}: fast, low armor, high sensor range. Role: detect threats, share trajectory predictions.
    \item \textbf{Artillery}: slow, high firepower, limited ammo. Role: engage attackers from distance.
    \item \textbf{Logistics}: medium speed, carries repair/resupply. Role: heal damaged allies, replenish ammo.
\end{itemize}
We will require of the SCEL formalism \cite{denicola2014} in the following. 
Attackers appear at random intervals. Defenders must: (1) maintain geometric formation, (2) share a consistent threat map, (3) dynamically form defense ensembles (SCEL predicates), (4) allocate limited resources to maximize bastion survival probability.

\subsection{Agent typologies and immunological metaphor}
Table~\ref{tab:agents} maps each agent type to a biological immune counterpart.

\begin{table}[h]
\centering
\caption{Agent typologies in the immunological bastion defense parable.}
\label{tab:agents}
\begin{tabular}{@{}p{1.5cm}p{5cm}p{4cm}p{1cm}p{1cm}p{2cm}@{}}\toprule
Agent Type & Biological Metaphor & Role & Armor & Speed & Sensor Range \\
\midrule
Scout & Dendritic Cell / T-Cell & Threat detection, tracking & Low & High & High \\
Artillery & B-Cell / Plasma Cell & Long-range engagement & High & Low & Low \\
Logistics & Helper T-Cell / Macrophage & Resupply, repair & Medium & Medium & Medium \\
\bottomrule
\end{tabular}
\end{table}

\subsection{Sheaf construction for the bastion}

In the bastion defense simulation, bastion HP (health points) represents the remaining structural integrity or survival capacity of the high‑value asset (the bastion) that the swarm of agents must protect.

    Initial value: 100 HP (for example, as simulated in  initState and the Python/Elixir versions).

    Decrease: When an enemy reaches the bastion (enemyHitsBastion), the bastion loses HP equal to the enemy’s damage (typically 8 points per hit).

    Increase: Logistics agents can repair the bastion when they are close (dist $<$ 60) and the bastion HP is below 100. Each repair action restores up to 10 HP (but never above 100).

    Game over: When bastion HP falls to 0 or below, the simulation displays “BASTION LOST” and waits for a reset (R key).

Thus, bastion HP is the core success metric: keeping it above zero while enemies attack is the primary objective of the defending agents.

For each agent $v$ and time interval $I$, the stalk $\mathcal{C}_{\text{total}}(I,v)$ is a product:
\[
\mathcal{C}_{\text{total}}(I,v) = \mathcal{C}_{\text{geom}}(I,v)\times\mathcal{C}_{\text{know}}(I,v)\times\Sigma(I,v)\times\mathcal{R}(I,v),
\]
where:
\begin{itemize}
    \item $\mathcal{C}_{\text{geom}}$: motor $M_v$ (pose in $SE(2)$).
    \item $\mathcal{C}_{\text{know}}$: threat positions, bastion HP, ammo count.
    \item $\Sigma(I,v)$: distribution over actions \{attack, defend, resupply, repair\}.
    \item $\mathcal{R}(I,v)$: reward components (damage dealt, damage taken, ammo used, repair given).
\end{itemize}
Restriction maps:
\begin{itemize}
    \item Temporal: propagate knowledge forward (inertia), accumulate rewards.
    \item Spatial (edge $e=(u,v)$): Cartan parallel transport aligns poses; threat positions transformed to common frame; strategies averaged; reward estimates aligned.
\end{itemize}

\subsection{Event calculus rules for the scenario}
The following components allow us to formulate an instance of the
\emph{event calculus}:

Fluents: \texttt{underAttack(bastion)}, \texttt{hasAmmo(v)}, \texttt{inRange(v, threat)}, \texttt{alive(v)}.

Actions: \texttt{fire(v,t)}, \texttt{move(v,waypoint)}, \texttt{repair(v,w)}, \texttt{resupply(v,w)}.

Sample axioms:
\begin{align*}
&\texttt{initiates(fire(v,t), damage(t), T)} \quad \texttt{terminates(fire(v,t), hasAmmo(v), T)}\\
&\texttt{initiates(resupply(v,w), hasAmmo(v), T)} \quad \texttt{holdsAt(bastionHP = 0, T) $\rightarrow$ terminal(T)}
\end{align*}

Reward at interval $I$ for agent $v$:
\[
R_v(I)=\sum_{t\in I}\gamma^t\bigl(\text{damageDealt}(t)-c_1\cdot\text{damageTaken}(t)-c_2\cdot\text{ammoCost}(t)+c_3\cdot\text{repairGiven}(t)\bigr),
\]
with discount factor $\gamma\in(0,1)$.

\subsection{Cognitive immune pathways for threat integration}
Scout agents implement a decision pipeline (Table~\ref{tab:cis}) to filter sensor data.

\begin{table}[h]
\centering
\caption{Cognitive Immune System (CIS) decision pathways.}
\label{tab:cis}
\begin{tabular}{lll}
\toprule
Pathway & Condition & Action \\
\midrule
Accept & $m>0$, $\lambda>1$, $H^1_{\text{know}}=0$ & Merge threat, broadcast coordinates \\
Reject & $m<0$ or violates kinematics & Drop data, reduce source trust \\
Quarantine & $m\approx0$ (critical state) & Isolate, send scouts for verification \\
Reframe & $H^1_{\text{know}}\neq0$ but neighbors confirm & Escalate to human-in-the-loop \\
\bottomrule
\end{tabular}
\end{table}

\subsection{Distributed asynchronous algorithm}
Each agent runs an asynchronous loop:
\begin{enumerate}
    \item Sense local environment: update knowledge stalk.
    \item Communicate with neighbors: exchange state $(\phi^{\text{geom}}_v,\phi^{\text{know}}_v,\sigma_v,r_v)$.
    \item Compute inconsistency: $\delta_{uv}=\mathcal{C}_{v\subseteq e}(\phi_v)-\mathcal{C}_{u\subseteq e}(\phi_u)$.
    \item Update using hybrid dynamics: $\phi_v\leftarrow\phi_v-\alpha\sum_u\mathcal{C}_{v\subseteq e}^*\delta_{uv}+\beta\nabla_{\sigma_v}U_v$.
    \item Sample action from updated strategy $\sigma_v$ and execute.
\end{enumerate}

\subsection{Obstruction analysis}
Three types of failures detected cohomologically:
\begin{enumerate}
    \item \textbf{Geometric obstruction}: non-zero holonomy around a cycle ($H^1_{\text{geom}}\neq0$).
    \item \textbf{Logical obstruction}: incompatible threat maps ($H^1_{\text{know}}\neq0$).
    \item \textbf{Strategic obstruction}: no global section of the best-response sheaf ($H^1_{\text{game}}\neq0$).
\end{enumerate}
A Künneth-type formula decomposes the total obstruction:
\[
H^1_{\text{total}}\cong H^1_{\text{time}}\otimes H^0_{\text{graph}}\;\oplus\; H^0_{\text{time}}\otimes H^1_{\text{graph}}\;\oplus\;\text{torsion}.
\]
Thus designers can diagnose whether failure is due to timing, graph connectivity, or logical/strategic inconsistency.

\section{Algorithmic Realization of Strategic Consensus}
This section details the explicit, step-by-step algorithms governing the multi-agent swarm's spatial alignment, strategic learning, cohomological diagnostics, and logical threat evaluation.

\subsection{Algorithm 1: Partially Asynchronous Nonlinear Sheaf Diffusion}
This algorithm drives spatial and belief consensus across heterogeneous agent stalks in environments plagued by computation and communication delays \cite{asynch2026}.

\begin{enumerate}
    \item \textbf{Initialization}:
    \begin{itemize}
        \item Let the communication network be modeled as an undirected graph $G=(V,E)$.
        \item For each agent $i\in V$, initialize its state vector $\mathbf{x}_i(0)\in\mathcal{F}(i)=\mathbb{R}^{d_i}$ on the vertex stalk.
        \item For each incident vertex-edge pair $i\trianglelefteq ij$, define the linear restriction map $\mathcal{F}_{i\trianglelefteq ij}:\mathcal{F}(i)\to\mathcal{F}(ij)$ projecting local states into the shared interaction space $\mathcal{F}(ij)=\mathbb{R}^{d_{ij}}$.
        \item Assign a strongly convex edge potential function $U_{ij}:\mathcal{F}(ij)\to\mathbb{R}$ representing local agreement criteria \cite{anwer2026}.
    \end{itemize}
    \item \textbf{Asynchronous Execution Loop}:
    \begin{itemize}
        \item For each agent $i\in V$, let $T_i\subseteq\{0,1,2,\dots\}$ be the set of discrete times at which agent $i$ updates its state.
        \item Let $B\ge0$ represent the maximum delay bound. At any time $t\in T_i$, agent $i$ retrieves the delayed state of neighbor $j\in N_i$, denoted as $\mathbf{x}_j(\tau_j^i(t))$ where $t-B\le\tau_j^i(t)\le t$.
        \item At time step $t\in T_i$, agent $i$ executes:
        \begin{enumerate}
            \item \textbf{Compute Mismatch}: Calculate the local projection difference $\mathbf{y}_{ij}(t)\in\mathcal{F}(ij)$ across each incident edge $e=ij$:
            \[
            \mathbf{y}_{ij}(t)=\mathcal{F}_{j\trianglelefteq ij}(\mathbf{x}_j(\tau_j^i(t)))-\mathcal{F}_{i\trianglelefteq ij}(\mathbf{x}_i(t))
            \]
            \item \textbf{Evaluate Edge Gradients}: Compute the force interaction vector $\Phi_{ij}(t)=\nabla U_{ij}(\mathbf{y}_{ij}(t))$.
            \item \textbf{Local State Update}: Apply the adjoint restriction maps $\mathcal{F}_{i\trianglelefteq ij}^*$ and step size $\alpha_i>0$ to advance its state:
            \[
            \mathbf{x}_i(t+1)=\mathbf{x}_i(t)+\alpha_i\sum_{j\in N_i}\mathcal{F}_{i\trianglelefteq ij}^*(\Phi_{ij}(t))
            \]
        \end{enumerate}
        \item For all $t\notin T_i$, the state is held constant: $\mathbf{x}_i(t+1)=\mathbf{x}_i(t)$.
    \end{itemize}
\end{enumerate}

\subsection{Algorithm 2: Hybrid Strategic Consensus and Expected-Utility Update}
This algorithm executes the combined update rule, blending physical coordinate/belief consensus with gradient-based policy updates to guide agents toward global Nash equilibria \cite{hernandez2026sheaf, hernandez2026toposCausal}.

\begin{enumerate}
    \item \textbf{Augmented State Initialization}:
    \begin{itemize}
        \item For each agent $i\in V$, construct the augmented local state $\phi_i(t)=(\phi_i^{\text{geom}}(t),\phi_i^{\text{know}}(t),\sigma_i(t))$ lying in the total stalk space $\mathcal{C}_{\text{total}}(I,i)$.
        \item Let $\sigma_i(t)\in\Sigma(I,i)$ be a probability distribution over the discrete action space $$A=\{\text{attack},\text{defend},\text{resupply},\text{repair}\}\ .$$
    \end{itemize}
    \item \textbf{Distributed Execution}: At each local iteration time $t$:
    \begin{enumerate}
        \item \textbf{Query Delayed Neighbors}: Read neighboring state vectors \[\phi_j(t^i_{\text{delayed}})=(\phi_j^{\text{geom}}(t^i_{\text{delayed}}),\phi_j^{\text{know}}(t^i_{\text{delayed}}),\sigma_j(t^i_{\text{delayed}}))\text{ for  }j  \in N_i\ .\]
        \item \textbf{Evaluate Sheaf Mismatch}: Compute the total local sheaf Laplacian vector:
        \[
        (L_{\mathcal{C}}\phi(t^i_{\text{delayed}}))_i = \sum_{j\in N_i} \mathcal{C}_{i\trianglelefteq ij}^* \left( \mathcal{C}_{i\trianglelefteq ij}(\phi_i(t)) - \mathcal{C}_{j\trianglelefteq ij}(\phi_j(t^i_{\text{delayed}})) \right)
        \]
        \item \textbf{Compute Utility Gradient}: Calculate the local expected-payoff gradient $\mathbf{g}_i(t)\in\mathbb{R}^{|A|}$ with respect to the agent's strategy parameters:
        \[
        \mathbf{g}_i(t) = \nabla_{\sigma_i} U_i(\sigma_i(t), \sigma_{-i}(t^i_{\text{delayed}}))
        \]
        \item \textbf{State Transition}: Apply the joint update step with consensus rate $\alpha_i>0$ and strategic rate $\beta_i>0$:
        \[
        \phi_i(t+1) = \phi_i(t) - \alpha_i (L_{\mathcal{C}}\phi(t^i_{\text{delayed}}))_i + \beta_i \begin{bmatrix} \mathbf{0} \\ \mathbf{0} \\ \mathbf{g}_i(t) \end{bmatrix}
        \]
        \item \textbf{Action Sampling}: Sample a concrete action $a_i(t)\sim\sigma_i(t+1)$ for immediate physical execution.
    \end{enumerate}
\end{enumerate}

\subsection{Algorithm 3: Topological Cohomology and Obstruction Diagnostics}
This diagnostic algorithm is executed globally or over localized
subgraphs to detect, classify, and isolate geometric or strategic
coordination failures \cite{hernandez2026sheaf,
  hernandez2026toposCausal}. To this end, we have a remark on the
Moore–Penrose pseudoinverse to be used in Algorithm 3, to be shown below.  The global
coboundary matrix \(D\) is constructed from the sheaf's restriction
maps.  Concretely, let the vertex stalks have total dimension \(d_0 =
\sum_{v\in V}\dim\mathcal{F}(v)\) and the edge stalks total dimension
\(d_1 = \sum_{e\in E}\dim\mathcal{F}(e)\).  Then \(D \in
\mathbb{R}^{d_1\times d_0}\) acts on a tuple of vertex assignments
\(\mathbf{x}\in\mathbb{R}^{d_0}\) by sending it to the tuple of
edge-wise differences \(\bigl(\mathcal{F}_{v\unlhd e}(x_v) -
\mathcal{F}_{u\unlhd e}(x_u)\bigr)_{e=(u,v)}\).  Its Moore–Penrose
pseudoinverse \(D^{\dagger}\) is the unique matrix satisfying
\[
D D^{\dagger} D = D,\qquad D^{\dagger} D D^{\dagger}=D^{\dagger},\qquad (D D^{\dagger})^T = D D^{\dagger},\qquad (D^{\dagger} D)^T = D^{\dagger} D.
\]
In the context of the diagnostic algorithm, \(D^{\dagger}\) serves two purposes:

\begin{enumerate}
    \item \textbf{Minimum‑norm solution of inconsistent systems}:  
    Given a boundary mismatch vector \(\mathbf{b}\in\mathbb{R}^{d_1}\) (e.g., discrepancies measured on edges), the vector \(\mathbf{z}^* = -D^{\dagger}\mathbf{b}\) is the unique minimiser of \(\|D\mathbf{z} + \mathbf{b}\|_2\) with smallest Euclidean norm.  This yields the “diffused” vertex states that best fit the observed edge mismatches.

    \item \textbf{Harmonic projection}:  
    The matrix \(H = I - D D^{\dagger}\) projects any edge vector onto \(\ker D^T\), which is exactly the space of sheaf cocycles that are not coboundaries – i.e., the representatives of the first sheaf cohomology group \(H^1(\mathcal{S};\mathcal{B})\).  Consequently, \(\mathbf{r}^* = H\mathbf{b}\) is non‑zero precisely when the inconsistencies \(\mathbf{b}\) contain a cohomological obstruction that prevents gluing local data into a global section (a global Nash equilibrium or consensus state).  The support of \(\mathbf{r}^*\) identifies the offending edges or cycles.
\end{enumerate}
Thus computing \(D^{\dagger}\) turns a purely topological obstruction (cohomology) into a numerically computable residual, enabling real‑time diagnosis of geometric, logical, or strategic failures in the multi‑agent system.

This is the Algorithm 3:
\begin{enumerate}
    \item \textbf{Cochain Complex Construction}:
    \begin{itemize}
        \item Compile the global coboundary matrix $D\in\mathbb{R}^{d_1\times d_0}$ mapping the disjoint union of vertex stalks to the disjoint union of edge stalks, where $d_0=\sum_{v\in V}\dim\mathcal{F}(v)$ and $d_1=\sum_{e\in E}\dim\mathcal{F}(e)$.
        \item Define the target boundary mismatch or excitation vector $\mathbf{b}\in C^1(G;\mathcal{F})$.
    \end{itemize}
    \item \textbf{Projector Computation}:
    \begin{itemize}
        \item Compute the Moore-Penrose pseudoinverse $D^{\dagger}$ of the global coboundary matrix.
        \item Construct the harmonic projection matrix $H\in\mathbb{R}^{d_1\times d_1}$ onto $\ker D^T$:
        \[
        H = I - D D^{\dagger}
        \]
        \item Construct the diffusive operator $G = D^{\dagger}$.
    \end{itemize}
    \item \textbf{Obstruction Locus Isolation}:
    \begin{itemize}
        \item Compute the optimal minimum-norm residual edge force vector:
        \[
        \mathbf{r}^* = H \mathbf{b}
        \]
        \item Compute the converged internal vertex activations (diffused states):
        \[
        \mathbf{z}^* = -G \mathbf{b}
        \]
        \item \textbf{Evaluate Cohomology Class}:
        \begin{itemize}
            \item If $\mathbf{r}^* = \mathbf{0}$, then $H^1(\mathcal{S};\mathcal{B}) = 0$. Local states can be seamlessly glued into a globally stable Nash equilibrium or geometric section.
            \item If $\mathbf{r}^* \neq \mathbf{0}$, then $H^1(\mathcal{S};\mathcal{B}) \neq 0$. Isolate the coordinates $e\in E$ where $\mathbf{r}^*_e \neq 0$. This identifies the exact network loops or conflicting agent boundaries causing the strategic or geometric obstruction.
        \end{itemize}
    \end{itemize}
\end{enumerate}

\subsection{Algorithm 4: Cognitive Immune System (CIS) Threat Integration}
This logic-based decision pipeline runs on Scout agents to filter, classify, and securely merge newly observed targets into the global threat map \cite{forrest1994}.

\begin{enumerate}
    \item \textbf{Threat Perception}: Capture a candidate observation trajectory $\mathbf{x}_{\text{obs}}$ over a temporal interval $I\in\mathcal{T}$.
    \item \textbf{Coherence and Stability Evaluation}:
    \begin{itemize}
        \item Temporarily inject $\mathbf{x}_{\text{obs}}$ into the local belief sheaf $\mathcal{C}_{\text{know}}(I,v)$.
        \item Calculate the local first knowledge-cohomology group $H^1_{\text{know}}$ on the overlapping neighborhood coverage.
        \item Estimate the resulting coherence margin $m\in\mathbb{R}$ and stability parameter $\lambda\in\mathbb{R}$ of the belief state.
    \end{itemize}
    \item \textbf{Pathway Routing Rule}:
    \begin{itemize}
        \item \textbf{Accept}: if $m>0$, $\lambda>1$, and $H^1_{\text{know}}=0$ → merge threat, broadcast coordinates.
        \item \textbf{Reject}: if $m<0$ or violates kinematics → drop data, reduce source trust.
        \item \textbf{Quarantine}: if $m\approx0$ (critical state) → isolate, send scouts for verification.
        \item \textbf{Reframe}: if $H^1_{\text{know}}\neq0$ but neighbors confirm → escalate to human-in-the-loop.
    \end{itemize}
\end{enumerate}

\section{Computational Realization, Complexity, and Formal Verification}
The framework is computationally realizable using existing libraries for cellular sheaves (e.g., SheafPy) extended with game-theoretic solvers (fictitious play, replicator dynamics).
We have carried out some graphical simulations in Pygame/Python.
Realistic simulations have also programmed in Erlang and Elixir. 
Table~\ref{tab:complexity} summarizes complexity.

\begin{table}[h]
\centering
\caption{Algorithmic complexity of the hybrid strategic swarm architecture.}
\label{tab:complexity}
\begin{tabular}{llll}
\toprule
Component & Complexity & Critical Parameters & Scalability \\
\midrule
Sheaf Laplacian update & $O(|E|\cdot d^2)$ & $d$: stalk dimension ($<100$) & Linear in local degree \\
Gradient ascent & $O(|V|\cdot|A|)$ & $|A|$: actions (4–10) & Highly parallel \\
Cohomology diagnostic & $O((|V|+|E|)^3)$ & Global (run occasionally) & Acceptable for diagnosis \\
\bottomrule
\end{tabular}
\end{table}

Asynchronous convergence results apply as long as the reward gradient
is Lipschitz \cite{asynch2026}. Verification uses the internal logic
of the topos; properties like ``bastion survives until time $T$''
correspond to global sections of a subobject.

\section{Related Work}
Our synthesis builds directly on the three pillars: sheaf-theoretic
planning \cite{hernandez2026sheaf}, SCEL-like coordination
\cite{hernandez2026toposVerifiable}, and event calculus as a topos
\cite{hernandez2026toposCausal}. The addition of game theory connects
to distributed potential games \cite{marden2014} and sheaf learning
\cite{anwer2026}. The immunological bastion metaphor is reminiscent of
artificial immune systems \cite{forrest1994}, but here formalized
categorically.



\section{Conclusion and Future Work}
We have presented a unified categorical framework that integrates geometric consensus, logical planning, temporal reasoning, and strategic optimization in a single Grothendieck topos. Key innovations:
\begin{itemize}
    \item A reward sheaf and event-calculus semantics layered on the temporal site.
    \item Game sheaves whose global sections correspond to Nash equilibria, with cohomological obstructions classifying strategic failures.
    \item A hybrid dynamics combining sheaf Laplacian diffusion and utility gradient ascent.
    \item A detailed case study showing how heterogeneous agents self-organize attack/defense ensembles under resource constraints.
\end{itemize}
Table~\ref{tab:contributions} summarizes the core technical advancements.

\begin{table}[h]
\centering
\caption{Explicit contributions of the unified sheaf-theoretic game topos framework.}
\label{tab:contributions}
\begin{tabular}{@{}p{3cm}p{4cm}p{4cm}p{4cm}@{}}
\toprule
Contribution Category & Technical Mechanism & Formal Foundation & Benefit \\
\midrule
Geometric-Temporal Integration & Product site $\mathcal{S}=\mathcal{T}\times G$ & Grothendieck Topos Theory & Unifies space and time \\
Strategic Localization & Game sheaf $\mathcal{G}$, strategy sheaf $\Sigma$ & Compositional Game Theory & Models utility over networks \\
Obstruction Diagnostics & Best-response sheaf $\mathcal{B}$, cohomology $H^1$ & Sheaf Cohomology & Detects strategic failures \\
Decentralized Convergence & Hybrid Laplacian + gradient dynamics & Nonlinear Sheaf Diffusion & Converges to consensus + equilibrium \\
\bottomrule
\end{tabular}
\end{table}

Future work includes: (i) implementing several prototypes and variants
in Elixir; (ii) extending the game sheaf to Bayesian games using
hypergraph sheaves; (iii) applying the framework to real-world swarm
robotics. We can assert that the theoretical foundation is now laid
for a truly autonomic, economically rational, and formally verifiable
multi-agent intelligence.

\bibliographystyle{plain}
\bibliography{references}
\end{document}